# Electronic structure and bond relaxation at Na/Ta(110) interfaces and 1D-chain and 2D-ring Ta metal structures on Na(110)

Maolin Bo,[a] Li Lei,[a] Chuang Yao,[a] Cheng Peng,[a]* Zhongkai Huang,[a] Chang Q. Sun[a,b]*,

[a]*Key Laboratory of Extraordinary Bond Engineering and Advanced Materials Technology (EBEAM) of Chongqing, Yangtze Normal University, Chongqing 408100, China*

[b]*NOVITAS, School of Electrical and Electronic Engineering, Nanyang Technological University, Singapore 639798, Singapore*

*E-mail: 20090008@yznu.cn; ecqsun@ntu.edu.sg

Abstract

We investigated the mechanism of Na/Ta(110) and Ta/Na(110) interfaces using a combination of bond–band–barrier (BBB) and zone-selective electron spectroscopy (ZES) correlation. We found that 7/9 ML and 8/9 ML Ta metal on a Na(110) surface form one-dimensional (1D) chain and two-dimensional (2D) ring structures, respectively. Moreover, we show that on Na(110), the Ta-induced Na(110) surface binding energy (BE) shifts are dominated by quantum entrapment. On the contrary, on a Ta(110) surface, the Na-induced Ta(110) surface BE shifts are dominated by polarization. Thus, the BBB and ZES strategy could potentially be used for designing 1D and 2D metals with desired structures and properties.

1. Introduction

Two-dimensional materials provide promising applications for a variety of optoelectronic devices owing to their unique structures and desirable properties[1-4]. However, based on their atomic configurations, most transition metals have unusual performances, for example, of their mechanical[5], electronic[6], catalytic[7], and magnetic[8] properties. Different electronic structures and atomic coordination environments of bulk and clean surface atoms are known to result in BE shifts in the core and valence bands of electrons. Core level shifts of surface atoms have been observed to change in a variety of defect generation[9-12,] alloy formation[13-16], and chemisorption[17-19] systems, with the changes generally being attributed to the transfer of charge. However, when the process of chemisorption is accompanied by surface reconstruction, which is often the case, the BE shifts are affected by the atomic bonding and the reconstruction-induced change in the surface atomic energy density and local strain.

In this paper, using zone-selective electron spectroscopy (ZES), we subtracted the electron spectroscopy results for Na(110) and Ta(110) surfaces before and after Ta and Na atom adsorption, respectively, and the residual X-ray photoelectron spectroscopy (XPS) and density-of-states (DOS) data retained only the features associated with the adsorption atoms of bond–band–barrier (BBB) correlation[20]. We found that Ta metal adsorption on an Na(110) surface results in the formation of one-dimensional (1D) chain and two-dimensional (2D) ring structures. Results confirmed our predictions that the bond energy ratio $\gamma$ can be directly related to bond identities ($d$, $E$) in Na/Ta(110) and Ta/Na(110) surfaces, which dictates the unusual behavior of solid surfaces and interfaces.

2. Principles

2.1 Bond-order-length-strength (BOLS) notation

According to the BOLS correlation[20], shorter and stronger bonds between under-coordinated atoms result in the BE shift of core-band electrons, which can be expressed as:

$$\begin{cases} \dfrac{d_i}{d_b} = c_i = \dfrac{2}{\{1+\exp[(12-z_i)/(8z_i)]\}} & \text{(local bond strain)} \\ \dfrac{E_i}{E_b} = c_i^{-m} \propto \left(\dfrac{E_v(I)-E_v(0)}{E_v(B)-E_v(0)}\right)^m = \gamma^m & \text{(local bond strength)} \end{cases}$$

(1)

where $i$ runs from the outermost atomic layer inward up to the third layer ($i \leq 3$); $z_i$ is the effective coordination number (CN) of an atom in the $i$th atomic layer of surface; $c_i$ is the coefficient of bond contraction; $d_i$ and $E_i$ are the bond length and bond energy in the $i$th atomic layer, respectively; $E_b$ is the bulk bond energy; and $d_b$ is the bulk bond length of the corresponding material. The bond nature indicator $m$ represents how the bond energy $E_i$ changes with the bond length $d_i$, and $m = 1$ for most metals.

## 2.2 Interface effect

In the interface region, if the atomic coordination number and the core-electron wavefunction change insignificantly, the bond energy uniquely determines the impurity-induced core level BE shifts:

$$\gamma = \frac{E_v(I)-E_v(0)}{E_v(B)-E_v(0)} = \frac{\Delta E_v(I)}{\Delta E_v(B)} = \frac{\Delta E_v(B)+\Delta E_v(i)}{\Delta E_v(B)}$$

(2)

Based on this expression, we can determine the interface bond strength with the given XPS decomposition information on the $v$th energy level of an isolated atom $E_v(0)$ and in the bulk $E_v(B)$. It is expected that both the peak intensities evolve upon interface or alloy formation; the intensity peak of the $v$th energy level of the interface $E_v(I)$ will increase, rendering a loss of intensity of the bulk $E_v(B)$ peak because the total number of electrons in a particular energy level is conserved.

One can drive the interface binding energy $E_v(I)$ and elucidate whether entrapment or polarization dictate the interface performance by ZES analysis of the electron spectroscopy. One can also calculate the adsorption-induced interface bond energy ratio with the known $\Delta E_v(B)$ reference derived from the surface density functional theory (DFT) calculation and XPS analysis.

$$\gamma = \frac{\Delta E_v(B) + \Delta E_v(i)}{\Delta E_v(B)} = E_I/E_b = \begin{cases} >1 & (T) \\ <1 & (P) \end{cases}$$

(3)

Here, $\gamma$ is the ratio of the bond energy in the interface region $I$ to that in the ideal constituent bulk $B$. If $\gamma > 1$, quantum entrapment $T$ dominates; otherwise, polarization occurs.

As a detectable quantity, the bond-energy ratio $\gamma$ can be directly related to bond identities ($d$, $E$). One can predict $\gamma$-resolved local bond strain $\varepsilon_l$ and relative bond energy density $\delta E_D$ of an alloy or interface by the following:

$$\begin{cases} \varepsilon_I = \frac{\Delta E_v(B)}{\Delta E_v(I)} - 1 = \gamma^{-1} - 1 & \text{(local bond strain)} \\ \delta E_D = (E_i/d_i^3)/(E_b/d_b^3) - 1 = (\frac{\Delta E_v(B)}{\Delta E_v(I)})^{-4} - 1 = \gamma^4 - 1 & \text{(bond energy density)} \end{cases}$$

(4)

**2.3 ZES and BBB correlation**

ZES[20] can distinguish the spectral features of electrons in the interfaces and surfaces. ZES proceeds by subtracting the electron spectroscopy reference from the DFT calculations or XPS experiments. The residual ZES spectrum is obtained by subtracting two electron spectra collected from a surface before and after the surface is chemically or physically conditioned, such as by defect generation and chemisorption, under the same measurement conditions. The reference electron spectrum is of a clean surface. ZES (i) distinguishes the spectral features of the atomic adsorption of the surface and the clean bulk, and removes the shared spectral features; (ii) purifies the spectral features owing to chemical or physical conditioning; (iii) retains the surface features owing to filtering out the information from the DFT calculations or XPS experiments.

BBB correlation[21] shows the behavior of atomic bonding, and the core- and valence-band electrons involved in the adsorption of atoms at the surface of the metal. This model shows that the valence band of the DOS has four features in the adsorption of metal surfaces: bonding electron pairs, nonbonding electron pairs, electronic holes, and antibonding dipoles. The validity of the BBB correlation mechanism has been confirmed from the geometrical structure and valence DOS as well as the bond-forming kinetics of adsorption on the metal surfaces. Furthermore, this technique can be used to statically and dynamically monitor surface processes such as defect generation, alloy formation, and chemical reactions.

## 2.4 DFT calculation methods

We calculate the BE shift distribution of the optimal Na/Ta(110) and Ta/Na(110) interfaces using first principles. The optimal geometric configurations of Ta(110) and Na(110) at different coverages are shown in the **Fig. 1**. The Vienna *ab initio* simulation package and the plane-wave pseudopotential are used in the calculations[22]. We also employed the Perdew–Burke–Ernzerhof exchange-correlation potentials[23]. The plane-wave cut-off is 400 eV. The Brillouin zone is calculated with special *k*-points generated in an 8 × 8 × 1 mesh grid. The Ta(110) and Na(110) surfaces are modeled with a periodicity of three layers. All surfaces have a vacuum space of 16 Å. One bottom layer is fixed in the bulk position and the other atomic layers are fully relaxed. In addition, the Na and Ta atoms adsorbed on the Ta(110) and Na(110) surfaces are allowed to be fully relaxed. The optimal atomic positions were determined by convergence of the total energy to within 0.01 meV.

## 3. Results and discussion

### 3.1 Ta 4*f* and Na 2*p* BE shift of Na/Ta(110) interface

ZES was invented to purify local and quantitative information on the bonding and electronic dynamics associated with Na adsorption on the Ta(110) surface. **Fig. 2a and b** shows the experimental ZES of the Ta 4*f* and Na 2*p* core level extracted by subtracting the spectra of the clean bulk from those collected at different coverages of Na atoms adsorption, such as 0.2 ML, 0.45 ML, and 0.6 ML adsorbed on the Ta(110) surface, after background correction and spectral area normalization[24]. The adsorption of Na atoms on the Ta(110) surface provided three contributions to the Ta 4*f* core band. The ZES result shows two valleys and one peak, which correspond to the bulk *B*, polarization *P*, and quantum entrapment *T*, respectively. The XPS experiment and DFT calculation showed the same results for the surface of a Ta 4*f* core band, as shown in **Fig. 2c**.

To extract quantitative information about the absolute value of the quantum entrapment from the XPS data, we need the values of $\Delta E_{4f}(B)$ for each constituent. Based on the measured surface dependence of the Ta 4*f* BE shifts, we extracted $\Delta E_{4f}(B)$ for 2.713 eV[25]. Using Eq. 3, we can calculate the *γ* values for quantum entrapment with the resolved $\Delta E_{4f}(i) = E_{4f}(i) - E_{4f}(B)$, as listed in **Table 1**. For Na monolayer adsorption at different coverages, 0.2 ML, 0.45 ML, and 0.6 ML of the Ta(110) surface, the bond energy ratio *γ* values are 1.053, 1.056, and 1.061 eV respectively. The corresponding core level shifts are 0.144, 0.151, and 0.166 eV respectively in comparison to the bulk value of 21.836 eV for a Ta 4*f* energy level. These results indicate that the monolayer Na adsorbed on the Ta(110) surface induced a Ta 4*f* level BE shift and quantum entrapment (*γ* > 1).

According to **Table 1**, the bond energy density of the Na-adsorbent Ta(110) surface should be lower than that of the ideal surface for the Na(110) skin. As the Na atom adsorption exhibits a higher work function than the Ta atoms, the Na$^{\delta+}$-Ta$^{\delta-}$ bond is entrapped with a positive charge on the Na atom and a negative Na 2$p$ BE shift, as is shown in **Table 2**. By comparing the computed energy shifts for differently reconstructed geometries of atomic bond length and atomic component with the BE shifts obtained from high-energy-resolution photoemission experimental measurements, we confirm that the local bond strain is formed by Na atoms, with coverages 0.2 ML and 0.6 ML, to the Ta (110) surface, yielding a shift from the atomic component of −0.288 and −0.568 eV, respectively, as is shown in the **Table 1**. Therefore, the Na 2$p$ electron polarization ($\gamma < 1$) and Ta 4$f$ quantum entrapment ($\gamma > 1$) represent reconstructed geometrical structures, with local bond strain and atomic components, respectively.

Using **Eq. 4**, we can calculate the bond energy densities $\delta E_D$ of the Na/Ta(110) interface, as shown in Table 1. These results demonstrate that the Ta(110) surface with Na-atom adsorption shows negative bond energy densities of Na 2$p$ electrons, which can be attributed to Na 2$p$ energy-level polarization and to the Ta(110) surface. Meanwhile, ZES obtained the Na adsorption-induced Ta 4$f$ level positive BE shift and bond energy densities. This indicates that Na adsorption induced the Ta local bond strain and chemical compound formation at the interface on the Ta(110) surface. Thus, the bulk shows negative intensities and surfaceexhibit positive intensities, while exhibiting BE shifts, as is shown in **Fig. 2**.

**3.2 Ta 4$f$ and Na 2$p$ BE shift of Ta/Na(110) interface**

To further confirm our predictions, we performed bond contraction on the binding energy and valence local DOS of the Na/Ta(110) interface with different coverages and structures based on the DFT calculations, as shown in the **Fig. 1** and **Fig. 3**. **Fig. 3** shows the core band BE shift of (a) Ta 4$f$ and (b) Na 2$p$, for clean Ta(110), 7/9 ML, and 8/9 ML Ta-atom adsorption on Na(110) calculated by DFT electron spectroscopy. It can be seen that the Ta 4*f* BE shifts from −34.318 eV for the 7/9 ML Ta-atom adsorption to −33.711 eV for 8/9 ML Ta-atom adsorption, while the Na 2*p* BE shifts from −23.854 eV for 7/9 ML Ta-atom adsorption to −23.985 eV for 8/9 ML Ta-atom adsorption; this illustrates the Ta 4*f* energy level is positive BE shift and the Na 2*p* energy level is negative BE shift. These results are in accordance with Hirshfeld population analysis which indicates that the Ta atom accepts electrons from the Na atom. It indicates that the interaction between the Ta atoms adsorbed on the Na(110) surface induces Ta 4*f* electron quantum entrapment and Na 2*p* electron polarization, and hence resulted in the positive or negative BE shifts and bond strain of Ta and Na levels, as shown in **Fig. 3** and **Table 1**. These DFT calculations further confirmed our BOLS predictions.

**3.3 Structure reconstruction of Na/Ta(110) and Ta/Na(110) interfaces**

We have addressed the problem of Na/Ta(110) interface electronic dynamics by measuring the energy shifts by XPS experiments and DFT calculations, as shown in **Fig. 1**. Researchers have been able to use photoemission spectroscopy to observe the changes of the energy shift and then calculate the change in bond lengths and geometric structure[26, 27]. There is now widespread support for these proposals, according to our recent study. Our calculations indicate that spontaneous bond contraction occurs at the Na/Ta(110) and Ta/Na(110) interfaces, as shown in **Table 2**. The DFT results show that the 1/3 ML Na-adsorbent Ta(110) surface bond contracts to 12.57% for the first atomic layers compared with the corresponding bulk values.

The experimental Na coverages 0.6 ML on Ta(110) 4*f* shows both the entrapment ($T$ = 21.989 eV) and the polarization states ($P$ = 21.522 eV). The $T$ value is below and the $P$ value is above the bulk value ($B$ = 21.836 eV), as shown in **Fig. 2a** and **Table 1**. The extents of quantum entrapment and polarization increase with the concentration of defects. Exercises provide not only a promising numerical approach for the quantitative information about the bond and electronic behavior but also consistent insight into the geometric structural dynamics of the Na/Ta(110) and Ta/Na(110) interfaces.

### 3.4 Bonding, nonbonding, antibonding, and electron-hole states of Na/Ta(110) and Ta/Na(110) interfaces

ZES shows valence band of bond-electron dynamics monitor surface and interface processes. The ZES profile was obtained from the DFT calculation of the Na/Ta(110) and Ta/Na(110) interfaces. Electron transfer appears in the valence electron band after monolayer Na adsorption. **Fig. 4** shows the valence DOS for the Na/Ta(110) and Ta/Na(110) interfaces at the most stable adsorption sites for different coverages. **Figs. 4a and b** show that the extent of the valence band BE shift depends on the Na and Ta atom coverage of Ta(110) and Na(110) respectively. Consistency between the predictions and observations revealed that the Na-atom adsorption enhances polarization and Ta-atom adsorption enhances entrapment, and the interaction of the Na-atom adsorption with Ta(110) skins makes Na atomic bonding weaker and Ta-atom adsorption with Na(110) skins makes Ta atomic bonding stronger.

Compared with and without Na-atom adsorption on the Ta(110) surface (**Fig. 5a**), Na-atom adsorption creates four additional DOS features: Na–Ta bonding states (from −4 to −6 eV), nonbonding electrons of Na–Ta (from −2 to −4 eV), $Na^{\delta+}$ polarized $Ta^{\delta-}$ atoms, $Na^{\delta+}$ electron-hole states (from 0 to -2 eV), and antibonding states (from 0 to +2 eV); these features are surprisingly consistent with the predictions of the BBB premise. Na–Ta electron-bonding proceeds in the following manner: the Ta atom accepts electrons from the neighboring Na atoms leaving behind $Na^{\delta+}$ electron-holes, Hirshfeld population analysis to record the charge transfer Ta from Na, and the Na *2p*-orbital polarization takes place with additional nonbonding electron states. The negative sign of the Ta atom means charge is gained from the Na atoms, as shown in **Table 2.**

The deformation charge density is shown in **Fig. 6**. We found that the formation of 1D chain and 2D ring structures of the Ta metal on the Na(110) surface is mainly due to changes in the electron distribution and Ta–Ta atomic bonding. In Fig. 6 are plotted the deformation charge densities of the 1D Ta metal chain (7/9 ML) and the 2D Ta metal ring (8/9 ML); the red charge density is negative, meaning that electrons are gathered in those parts of the structure, and the blue charge density is positive, meaning that electrons are lost from those parts.

**4. Conclusion**

Incorporating ZES and BBB correlations with DFT calculations and XPS measurements has led to consistent insight into the physical origin of the localized edge states of Na/Ta(110) and Ta/Na(110) interfaces. The detailed work and results achieved are as follows:

1. We systematically examined Na/Ta(110) and Ta/Na(110) interfaces using a combination of photoelectron spectrometric analysis and DFT calculations. Consistency between the predictions and observations revealed that under-coordinated atoms may be attributed to the transfer of charge, local bond strain, and bond energy density, which results in BE shifts.

2. Based on the ZES method, we analyzed the Na/Ta(110) and Ta/Na(110) interfaces. Consistency between predictions and observations revealed that Na-atom adsorption enhances polarization and Ta-atom adsorption enhances entrapment. Owing to bond formation, the Ta atoms squeeze the top layer because the Ta-Ta bonds of the skins are much stronger than the Na-Ta bonds, and cause surface reconstruction, which results in the formation of atomic chain and ring structures of the Ta metal.

3. Based on BBB correlation, the results showed the formation of interface chemical bonds with four additional DOS features: bonding electrons, nonbonding electrons, electronic holes, and antibonding dipoles.

The concepts of atomic-adsorption-induced quantum entrapment and polarization are essential for understanding the bonding and electronic behavior of under- and hetero-coordinated atoms at interfaces or nearby impurities, which is beyond the scope of the available approaches.

**Acknowledgment**

Financial support was provided by the NSF (Grant No. 11747005), the Science and Technology Research Program of Chongqing Municipal Education Commission (Grant No. KJ1712299), and Yangtze Normal University (Grant No. 2016XJQN28 and 2016KYQD11).

**Table and figure captions:**

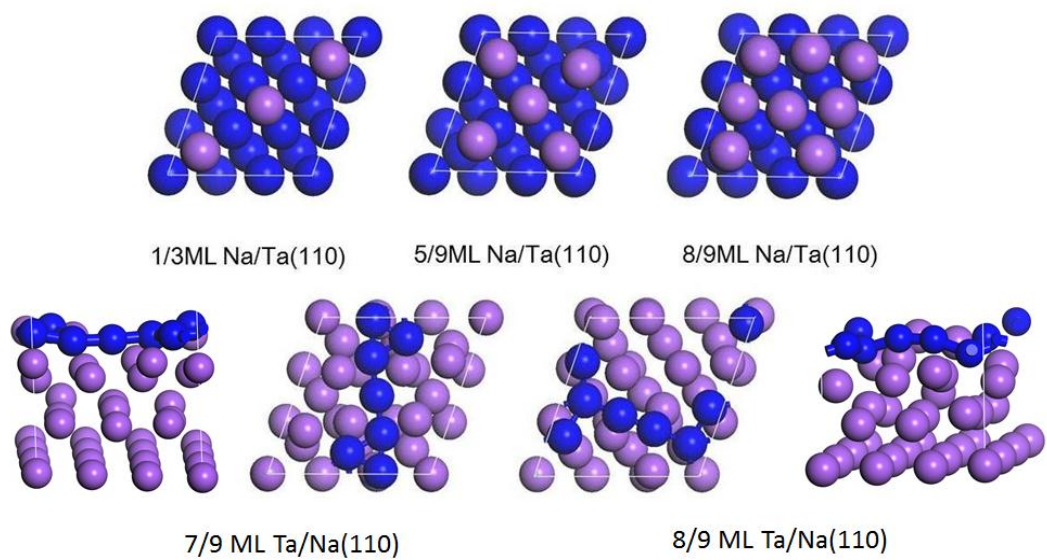

**Fig. 1** Geometric configurations of Na/Ta(110) and Ta/Na(110) interfaces at different coverages. Violet and blue correspond to Na and Ta atoms respectively.

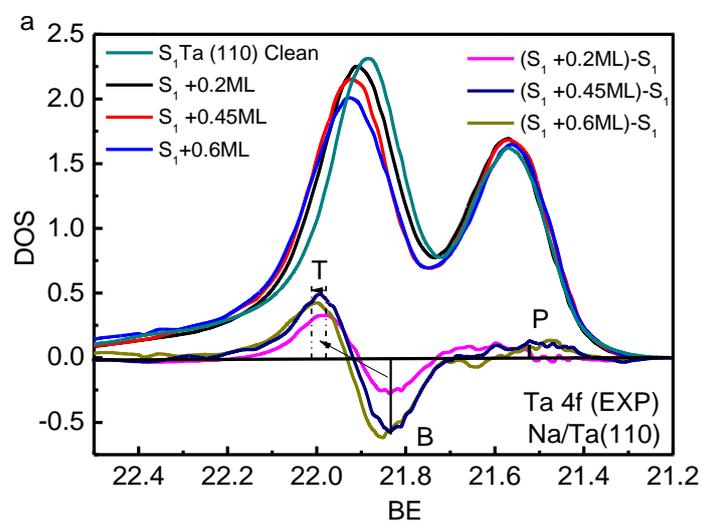

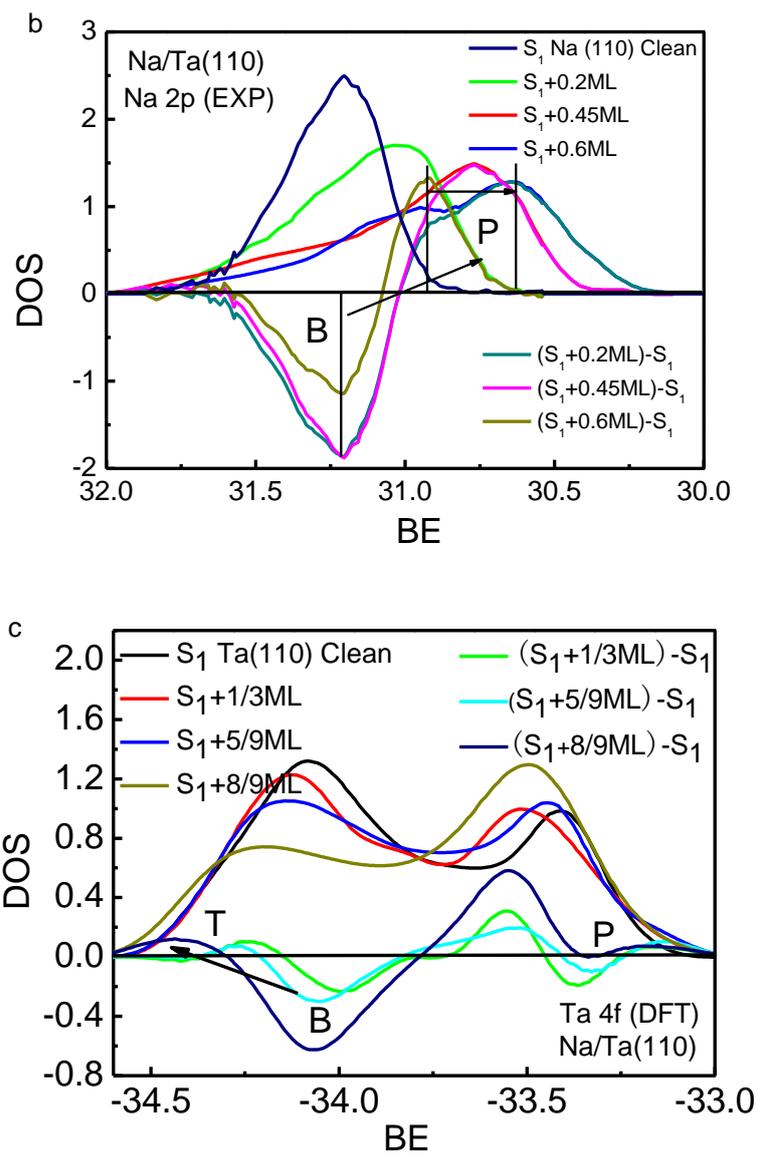

**Fig. 2** ZES of BE shifts of Na/Ta(110) interface at different coverages[24]. The results show the spectroscopic features of the bulk *B*, polarization *P*, and quantum entrapment *T* of the core band.

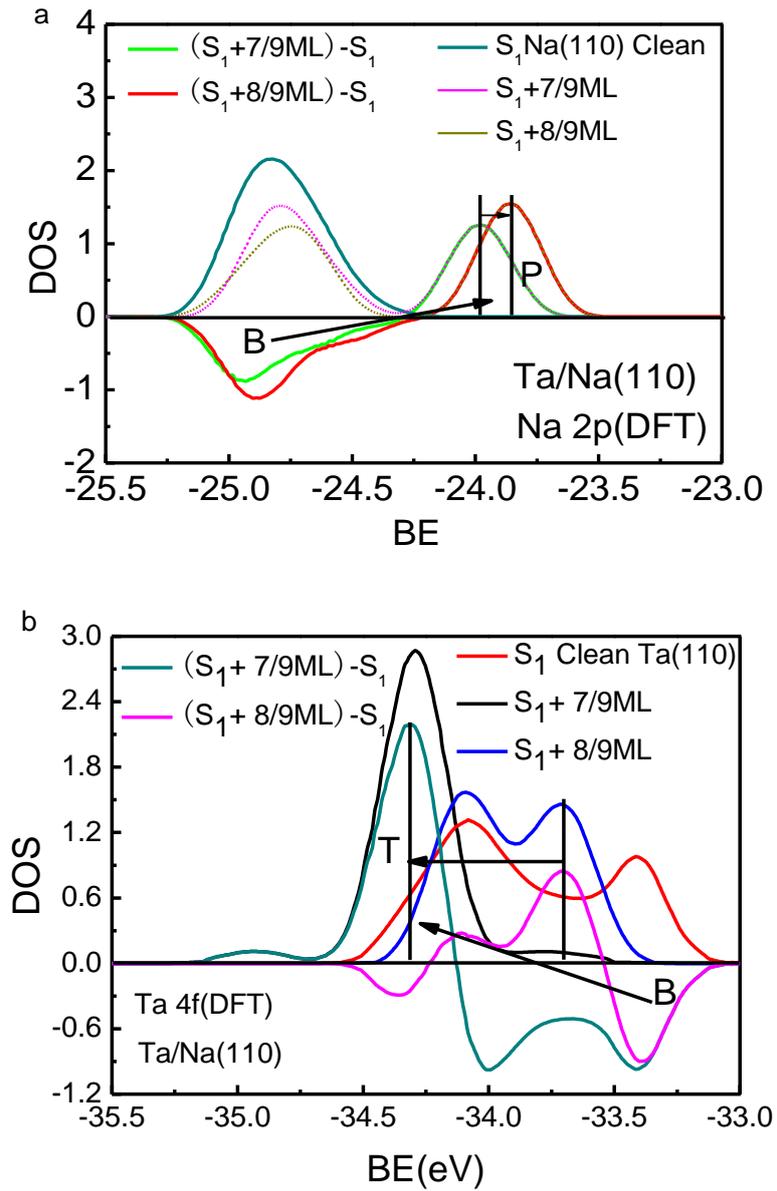

**Fig. 3** ZES of BE shifts of Ta/Na(110) interface at different coverages. The results show the spectroscopy features of the bulk *B*, polarization *P*, and quantum entrapment *T* of the core band.

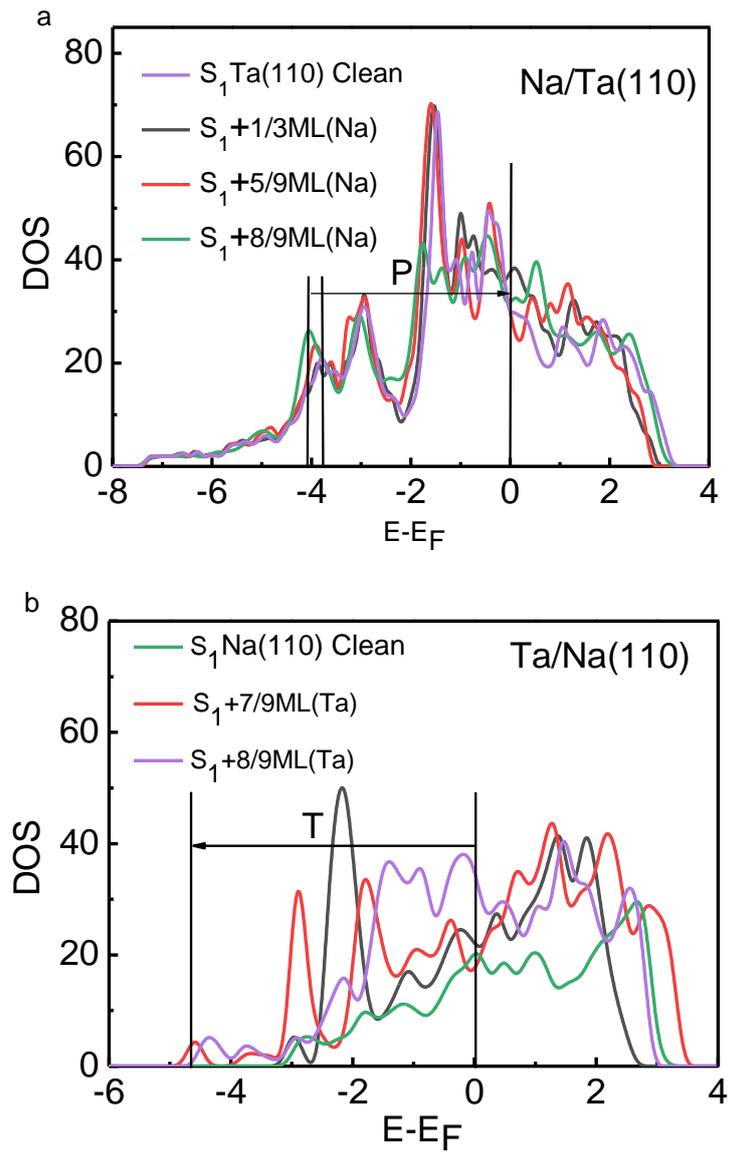

**Fig. 4** Valence band BE Shifts of Na/Ta(110) and Ta/Na(110) interfaces at different coverages.

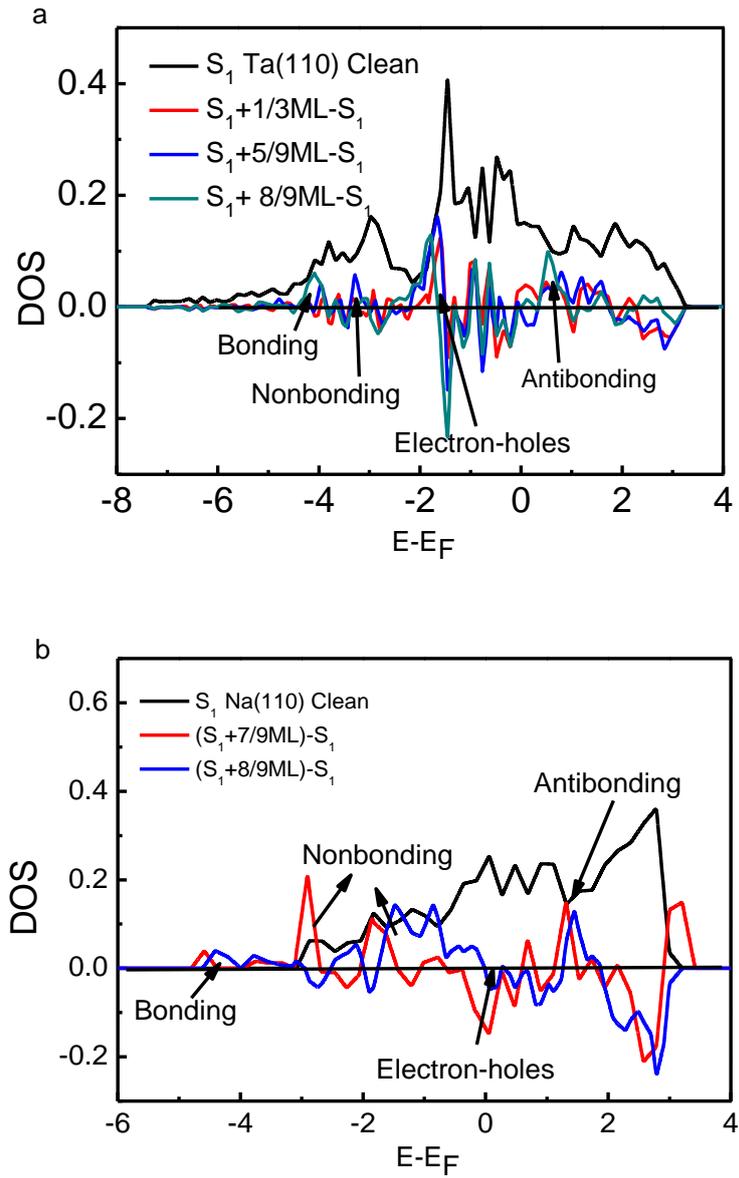

**Fig. 5** Coverage dependence of ZES of Na/Ta(110) and Ta/Na(110) interfaces. The profiles exhibit four valence DOS features: antibonding, electron-hole, nonbonding, and bonding states.

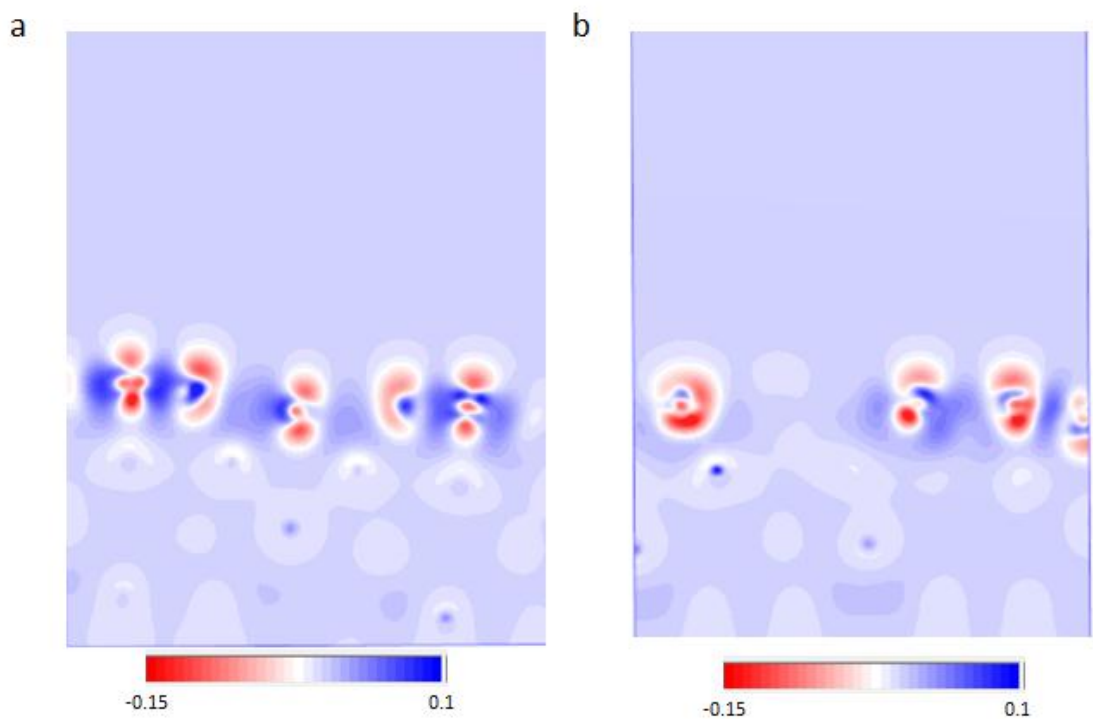

**Fig. 6** Deformation charge densities of structures of (a) 7/9 ML and (b) 8/9 ML Ta on an Na(110) surface.

**Table 1** BOLS derived BE shifts $\Delta E_v(i) = \Delta E_v(I) - \Delta E_v(B)$[25, 28], the bond energy ratio $\gamma$(%), local bond strain $-\varepsilon_l$(%), relative bond energy density $\delta E_D$(%) at different coverages of Na/Ta(110) interfaces.

| Experiment[24] | Adsorption concentration(L) | $E_v(i)$ | $\Delta E_v(i)$ | $\gamma$ | $\varepsilon_l$ | $\delta E_D$ |
|---|---|---|---|---|---|---|
| Bulk | 0 | 21.836 | 0 | 1 | 0 | 0 |
| Ta 4$f$ | 0.2 ML | 21.980 | 0.144 | 1.053 | -5.040 | 22.982 |
|  | 0.45 ML | 21.987 | 0.151 | 1.056 | -5.272 | 24.192 |
|  | 0.6 ML | 22.002 | 0.166 | 1.061 | -5.766 | 26.814 |
| Bulk | 0 | 31.211 | 0 | 1 | 0 | 0 |
| Na 2$p$ | 0.2 ML | 30.923 | -0.288 | 0.880 | 13.630 | -40.017 |
|  | 0.45 ML | 30.773 | -0.438 | 0.816 | 22.312 | -55.320 |
|  | 0.6 ML | 30.643 | -0.568 | 0.763 | 30.987 | -66.031 |

**Table 2** Bond contraction ratio, work function, and charge transfer for different coverages of Ta/Na(110) and Na/Ta(110) interfaces.

| DFT | Adsorption concentration (ML) | $d_i$ (Å) | Bond contraction ratio | Work function (eV) | Charge [c] |
|---|---|---|---|---|---|
| Na/Ta(110) | 1/3 | 3.249 [a] | 12.57% | 2.466 | 0.69(Na) |
|  | 5/9 | 3.185 [a] | 12.86% | 3.202 | 0.57(Na) |
|  | 8/9 | 3.294 [a] | 11.36% | 3.626 | 0.44(Na) |
| Ta/Na(110) | 7/9 | 3.050 [a], 2.572 [b] | 17.92%, 10.07% | 3.202 | -0.66(Ta) |
|  | 8/9 | 3.038 [a], 2.507 [b] | 18.24%, 12.34% | 3.138 | -0.72(Ta) |

[a]The Ta−Na bond length $d_i$ represents the distance of the first and second atomic layers, the initial Na-Na chemical bond is 3.716Å. [b]The Ta−Ta bond length $d_i$ represents distance of the first atomic layer, the initial Ta-Ta chemical bond is 2.860Å. [c]Negative sign indicates charge gain, otherwise, a charge loss occurs.

**Graphical abstract**

1. The geometric structure is one dimensional atomic chain of Ta Metal.

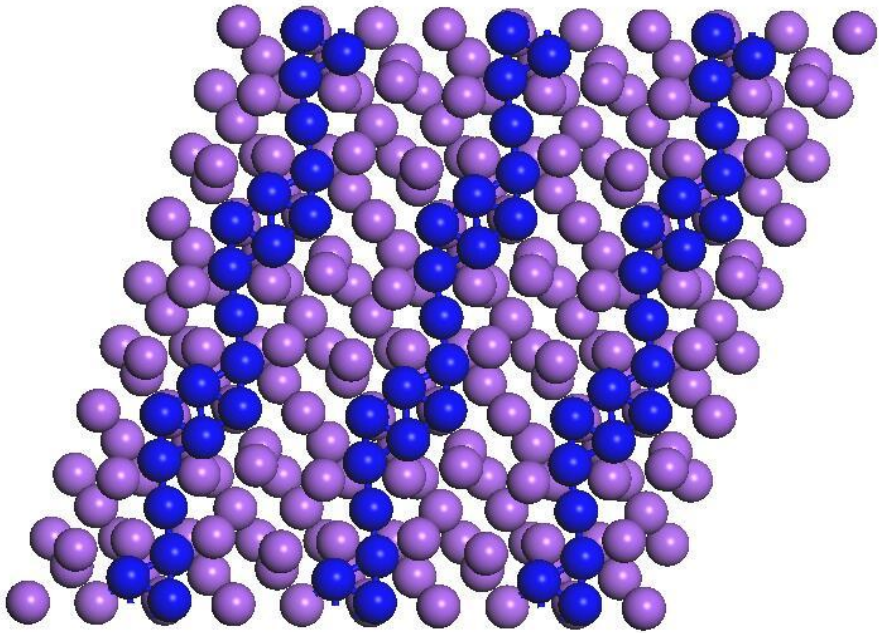

2. The geometric structure is Two dimensional ring of Ta metal.

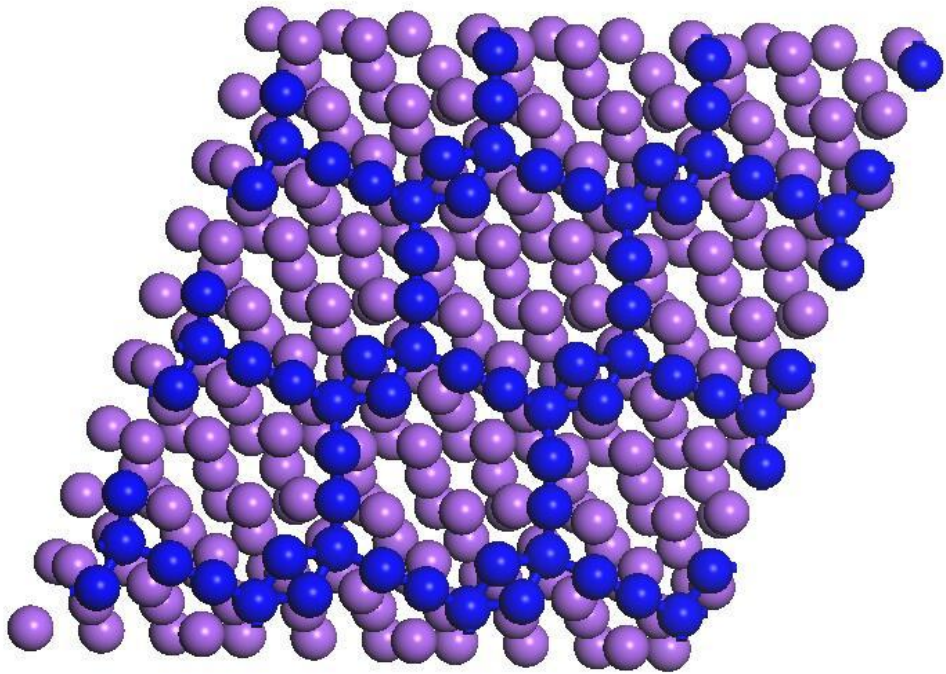